\documentclass[aps,physrev,twocolumn,superscriptaddress]{revtex4-2}

\usepackage{amsmath, amsthm, amssymb, mathtools}

\newcommand{\mybar}[3]{%
    \mathrlap{\hspace{#2}\overline{\scalebox{#1}[1]{\phantom{\ensuremath{#3}}}}}\ensuremath{#3}
}

\newcommand{\altbar}[1]{%
    \mybar{0.8}{0.6pt}{#1}
}

\newcommand{\inner}[2]{\langle#1, #2 \rangle} 

\begin{document}
\title{Impedance in Periodically Driven Stochastic Systems}

\author{Bart Wijns}
\affiliation{UHasselt, Faculty of Sciences, Theory Lab, Agoralaan, 3590 Diepenbeek, Belgium}
\author{Branko Meeus}
\affiliation{UHasselt, Faculty of Sciences, Theory Lab, Agoralaan, 3590 Diepenbeek, Belgium}
\address{Data Science Institute, Faculty of Sciences, UHasselt—Hasselt University, Agoralaan, 3590 Diepenbeek, Belgium}
\author{Jef Hooyberghs}
\affiliation{UHasselt, Faculty of Sciences, Theory Lab, Agoralaan, 3590 Diepenbeek, Belgium}
\address{Data Science Institute, Faculty of Sciences, UHasselt—Hasselt University, Agoralaan, 3590 Diepenbeek, Belgium}
\author{Bart Cleuren}
\email[]{bart.cleuren@uhasselt.be}
\affiliation{UHasselt, Faculty of Sciences, Theory Lab, Agoralaan, 3590 Diepenbeek, Belgium}
\date{\today }
\begin{abstract}
We study the time-dependent currents arising in periodically driven stochastic systems. In the linear regime of small driving, a closed expression is obtained for the impedance/admittance associated with these currents. This expression leads directly to an interpretation in terms of an equivalent electrical circuit. For a stochastic system with $N$ states, the electrical circuit consists of precisely $N$ parallel branches, with each branch comprising a resistor and a capacitor in series. The low- and high-frequency limits of these currents are calculated, and the results are generalized to more general current expressions. We demonstrate our findings across several archetypal settings, illustrating the potential to detect specific broken symmetries or the graph topology associated with the stochastic process.
\end{abstract}
\maketitle
%%%% SECTION 1 %%%%%%%%%%%%
\section{Introduction}
Transport phenomena are ubiquitous in nature, with examples ranging from macroscopic electrical currents to cellular molecular motors. The presence of an external driving force in these examples induces the transport. For small external driving, the governing transport equations can be linearized, yielding well-known results such as Fourier's and Fick's laws, the Green-Kubo relations, and the fluctuation-dissipation theorem~\cite{Green1954,Kubo1957,Hayashi2006,Marconi2008}. Macroscopic flows in this regime are therefore well-known results~\cite{ishii1985}. On the microscopic side, the theory of stochastic thermodynamics was developed to study (fluctuating) thermodynamic quantities related to small-scale systems~\cite{Seifert2008, Seifert2012, VandenBroeck2015, Peliti2021}. Recent work considered periodically driven systems 
~\cite{brandner2015thermodynamics,proesmans2015onsager,brandner2016periodic,Proesmans2016,Proesmans2019}, again with a focus on the linear response regime. In these works, thermodynamic quantities such as heat, work, and entropy (production) were obtained as time averages over one period of the driving, thereby losing the time variations within that period. In~\cite{Cleuren2019}, a fully time-dependent result was obtained for stochastic particle transport through a single-level quantum dot, including a definition of stochastic impedance. In~\cite{Forastiere2022}, full-time results were obtained with a focus on thermodynamic properties of these systems. This paper introduces the definition of stochastic impedance for arbitrary finite-state stochastic systems, with a focus on probability currents and potential applications.

The outline of this paper is as follows: Section~\ref{sec:model} introduces the generic finite-state stochastic system in which we will calculate the microscopic currents. The analogy with electrical currents and impedance is given in Section~\ref{sec:circuit-analogy}. Sections~\ref{sec:low-frequency} and~\ref{sec:high-frequency} explore the low and high frequency limits, respectively, of the result. In Section~\ref{sec:composite}, we introduce an expression to study more composite currents. We conclude in Section~\ref{sec:example} with an example of the method applied to specific example systems. 

%%%% SECTION 2 %%%%%%%%%%%%%%%%%%%%%%%%%%%%%%
\section{Periodically driven stochastic systems}\label{sec:model}
The starting point of our discussion is a generic stochastic system with $N$ discrete states. A master equation describes the time evolution caused by the transitions between these states
\begin{equation}\label{master-eqn}
\dot{\mathbf{P}}(t) = \mathbf{W}(t) \mathbf{P}(t).
\end{equation}
with $\mathbf{P}(t)$ the probability vector with components $P_n(t)$ $(n \in \{1,\ldots,N\})$ the probability for the system to be in state $n$ at time $t$. $\mathbf{W}(t)$ is a stochastic $N\times N$ matrix whose elements $W_{mn}(t)$ $(m\neq n)$ are the transition rates to go from state $n$ to state $m$ at time $t$ and $W_{nn}(t)=-\sum_{m\neq n}W_{mn}(t)$. As we want to keep the discussion general, these rates are assumed to satisfy the following two minimal conditions:
\begin{enumerate}
    \item[(i)] In the absence of an external driving, the system evolves towards a unique state of thermal equilibrium. Mathematically, this is equivalent to an irreducible system that satisfies the detailed balance condition.\label{cond_1}
    \item[(ii)] For those transitions which are affected by the driving, the rates depend continuously on the argument $F g(t)$ with $F$ a measure of the strength of the external driving, and $g(t)$ an arbitrary periodic function.
\end{enumerate}
Our focus will be on the time dependence of the probability currents
\begin{equation}\label{general-current}
J_{mn}(t)=W_{mn}(t)P_n(t)-W_{nm}(t)P_m(t).
\end{equation}
We consider from now on the long-time limit, i.e.\ the state of the system after all transients have vanished. In this limit, both the solution to Eq.~(\ref{master-eqn}) and the currents become periodic in time.\\
\begin{figure}[t]
\includegraphics[width=0.9\columnwidth]{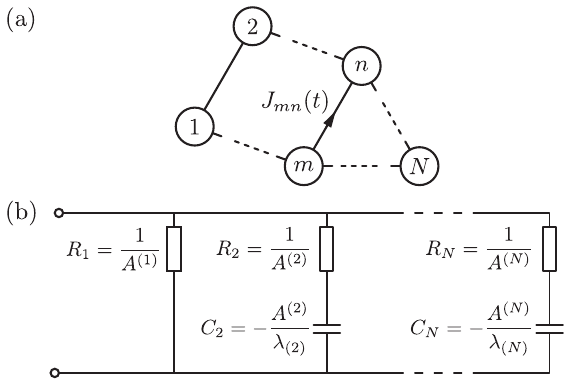}
\caption{(a) Diagram of the $N$ states of the stochastic system. The solid lines represent transitions between states. (b) The equivalent electrical circuit whose impedance has the same form as (\ref{simple-impedance}). The indices $mn$ on the 
coefficients $A^{(k)}_{mn}$ have been dropped for notational simplicity. The dotted lines represent multiple branches whose structure is 
identical to the drawn branches.}\label{fig:setup_circuit}
\end{figure}
Calculating the currents in Eq.~(\ref{general-current}) is a hopeless task for general time-dependent rates and system size $N$. However, as we show below, analytical expressions can be obtained in the regime of \emph{small} external driving, where linear response holds. Given the two minimal conditions, the first-order approximation of the transition matrix follows immediately:
\begin{equation}\label{w-expansion}
\mathbf{W}(t) \approx \mathbf{W}^{\text{eq}} + \altbar{\mathbf{W}} F g(t).
\end{equation}
The equilibrium matrix $\mathbf{W}^{\text{eq}}$ serves to define the (unique) equilibrium state $\mathbf{P}^{\text{eq}}$ as the solution of $\mathbf{W}^{\text{eq}} \mathbf{P}^{\text{eq}} = 0$. The matrix $\altbar{\mathbf{W}}$ is the derivative of $\mathbf{W}(t)$ w.r.t. $F$, evaluated at $F=0$. Following the methods used in~\cite{Proesmans2016, Proesmans2019}, the first-order time-periodic solution of the master equation can be expressed solely in terms of the adiabatic solution $\mathbf{P}^{\text{ad}} (t)$, defined as the solution to $\mathbf{W} (t) \mathbf{P}^{\text{ad}} (t) = 0$, yielding
\begin{equation}
\mathbf{P}(t) \approx \mathbf{P}^{\text{ad}} (t) - \int \limits_0^{\infty} d \tau e^{\mathbf{W}^{\text{eq}} \; \tau} \; \dot{\mathbf{P}}^{\text{ad}} (t-\tau) \label{approx-prob}.
\end{equation}
We can expand this further by evaluating the matrix exponential $e^{\mathbf{W}^{\text{eq}} \, \tau}$ and then performing the matrix product. Since $\mathbf{W}^{\text{eq}}$ is diagonalizable~\cite{vankampen}, it has a full set of real eigenvalues $\lambda_k$ and eigenvectors $\mathbf{v}_k$, with $k \in \{1, \dots , N\}$. Furthermore, all eigenvalues are negative except for one, which is zero. The associated eigenvector of the zero eigenvalue is the equilibrium solution ($\lambda_1 = 0$ and $\mathbf{v}_1 = \mathbf{P}^{\text{eq}}$). Considering the adiabatic solution to be a first-order perturbation of this equilibrium solution,
\begin{equation}
   \mathbf{P}^{\text{ad}} (t) \approx  \mathbf{P}^{\text{eq}} + \altbar{\mathbf{P}}^{\text{ad}} Fg(t),
\end{equation}
and carrying out the diagonalization (details can be found in appendix~\ref{app:Diag}), we find 
\begin{equation}\label{p_approx}
\mathbf{P}(t) \approx \mathbf{P}^{\text{ad}} (t) - \int \limits_0^{\infty} d \tau \sum \limits_{k = 2}^N \mathbf{v}_k e^{\lambda_k \, \tau} \langle 
\mathbf{v}_k , \altbar{\mathbf{P}}^{\text{ad}} \rangle F \dot{g}(t-\tau).
\end{equation}
In the above, the inner product is defined as
\begin{equation}\label{inner}
\langle \mathbf{v} , \mathbf{w} \rangle = \sum \limits_{n = 1}^N \frac{v_n w_n}{P^{\text{eq}}_n}.
\end{equation}
Note that $\langle \mathbf{v}_1 , \altbar{\mathbf{P}}^{\text{ad}} \rangle = 0$ due to orthogonality. Condition (i) guarantees that the matrix $\mathbf{W}^{\text{eq}}$ is 
self-adjoint with respect to this inner product~\cite{vankampen}. Hence the eigenvectors are orthonormal, so $\langle \mathbf{v}_k , 
\mathbf{v}_l \rangle = \delta_{kl}$.\\

Turning back to the currents Eq.~(\ref{general-current}), we note that the right-hand side of Eq.~(\ref{p_approx}) can be interpreted as an adiabatic part $\mathbf{P}^{\text{ad}}(t)$ and a non-adiabatic part $\mathbf{P}^{\text{nad}} (t)$. Correspondingly, the current can also be split as such,
\begin{equation}
J_{mn}(t) = J_{mn}^{\text{ad}}(t) + J_{mn}^{\text{nad}}(t).
\end{equation}
For the adiabatic part, we find up to first order in $F$,
\begin{eqnarray}
J_{mn}^{\text{ad}} (t) &=& W_{mn} (t) P^{\text{ad}}_n (t) - W_{nm} (t) P^{\text{ad}}_m (t) \nonumber \\
& \approx & F g(t) \big(W_{mn}^{\text{eq}} \altbar{P}_n^{\text{ad}} + \altbar{W}_{mn} P_n^{\text{eq}} \nonumber \\
&& \hspace{1.4cm}-  W_{nm}^{\text{eq}} 
\altbar{P}_m^{\text{ad}} - \altbar{W}_{nm}P_m^{\text{eq}} \big).\label{current_ad}
\end{eqnarray}
The non-adiabatic current reduces to
\begin{eqnarray}\label{current_nad}
J_{mn}^{\text{nad}} (t) &=& -\sum \limits_{k = 2}^{N} \inner{\mathbf{v}_k}{\altbar{\mathbf{P}}^{\text{ad}}} \Big(W_{mn}^{\text{eq}} v_{k, n} - W_{nm}^{\text{eq}} 
v_{k, m} \Big)\nonumber \\
&& \hspace{0.8cm} \times \int \limits_{0}^{\infty} d \tau e^{\lambda_k \tau} F \dot{g} (t - \tau),
\end{eqnarray}
At this point in the discussion, having obtained the general time-periodic first-order expression for the currents, it is illuminating to adopt a complex notation, as is done, for example, in the context of electrical circuits driven by an alternating voltage. Such a notation allows combining amplitude and phase into a single complex quantity. For the driving and current, we write
\begin{equation}
    Fg(t)=\mbox{Re}\left[\mathbf{F}e^{i \omega t} \right] \;\;\;\; ; \;\;\;\; J_{mn}(t)=\mbox{Re}\left[\mathbf{J}_{mn}e^{i \omega t} 
    \right]
\end{equation}
The impedance $\mathbf{Z}_{mn}$ is then determined by the ratio 
\begin{equation}\label{impedance}
    \frac{1}{\mathbf{Z}_{mn}}\equiv \frac{\mathbf{J}_{mn}}{\mathbf{F}}=\sum \limits_{k = 1}^N \frac{A^{(k)}_{mn} i \omega}{i \omega - 
    \lambda_k},
\end{equation}
with coefficients $A^{(k)}_{mn}$
\begin{align}\label{coeffs-simple}
&A_{mn}^{(1)} = W_{mn}^{\text{eq}} \altbar{P}_n^{\text{ad}} + \altbar{W}_{mn} P_n^{\text{eq}} -  W_{nm}^{\text{eq}} \altbar{P}_m^{\text{ad}} - \altbar{W}_{nm}P_m^{\text{eq}} \nonumber \\
&A_{mn}^{(k>2)}=-\inner{\mathbf{v}_k}{\altbar{\mathbf{P}}^{\text{ad}}} \left(W_{mn}^{\text{eq}} v_{k, n} - W_{nm}^{\text{eq}} v_{k, m} \right).
\end{align}
Eq.~\eqref{impedance} is the central result of this work. Note that while the coefficients $A^{(k)}_{mn}$ are expressed in terms of the 
eigenvectors of $\mathbf{W}^{\text{eq}}$, they are independent of the choice of eigenbasis. A proof is given in appendix~\ref{app:degens}.
%%%% SECTION 3 %%%%%%%%%%%%%%%%%%%%%%%%
\section{Electrical circuit analogy}\label{sec:circuit-analogy}
Equations~\eqref{current_ad} and~\eqref{current_nad} give a linear relationship between the current and the driving force. A similar relationship holds in electrical networks, where Ohm's law applies. For electrical systems with alternating voltages, one defines 
the impedance $Z$ as
\begin{equation}
V = ZI,
\end{equation}
where $V$ is the applied voltage and $I$ is the current through the network. The analogy with equation~\eqref{impedance} is evident. Furthermore, we can compare this to the impedance addition law for parallel circuits, which states that the total impedance of $n$ circuits in parallel is given by
\begin{equation}
\frac{1}{\mathbf{Z}} = \sum \limits_{k = 1}^n \frac{1}{\mathbf{Z}_k},
\end{equation}
where $Z_k$ is the impedance of the $k-$th individual component. We can rewrite the impedance in~\eqref{impedance} as
\begin{equation}
\frac{1}{\mathbf{Z}_{mn}} = \sum \limits_{k = 1}^N \frac{A^{(k)}_{mn} i \omega}{i \omega - \lambda_k} \label{simple-impedance}
= \sum_{k = 1}^N {\left( \frac{1}{A^{(k)}_{mn}} - \frac{\lambda_k}{i \omega A^{(k)}_{mn}} \right)}^{-1},
\end{equation}
which can be interpreted as the impedance of a parallel circuit consisting of $N$ components with individual impedances
\begin{equation}
\mathbf{Z}_{mn, k} = \frac{1}{A^{(k)}_{mn}} - \frac{\lambda_k}{i \omega A^{(k)}_{mn}}.
\end{equation}
These can be interpreted as the impedance of a device consisting of a resistor $R_{mn, k} = \frac{1}{A^{(k)}_{mn}}$ and a capacitor $C_{mn, k} = -\frac{A^{(k)}_{mn}}{\lambda_k}$, in series. Hence, the full circuit can be drawn as shown in Figure~\ref{fig:setup_circuit}. The electrical analogy is not perfect, since this only yields a physically possible circuit if $A^{(k)}_{mn} > 0$ for all $k$. This positivity condition will generally not hold, and as a consequence, the probability current can exhibit much richer behavior than an electrical current.

In section~\ref{sec:high-frequency}, we will see that for undriven transitions, the impedance diverges as $\omega$ tends to infinity. Therefore, it is more natural to consider the inverse of the impedance, the conductance $\sigma_{mn}(\omega)$. From~\eqref{impedance} it follows that this is a bounded function of $\omega$ for any transition in any finite system. The conductivity of a branch $k \geq 2$ for the transition $n \to m$ is then
\begin{equation}
    \sigma_{mn,k}(\omega) = \frac{i \omega C_{mn,k}}{i \omega C_{mn,k} R_{mn,k} + 1}.
\end{equation}

%%%% SECTION 4 %%%%%%%%%%%%%%%%%%%%%%%%
\section{The low frequency limit}\label{sec:low-frequency}
We investigate the behavior of the impedance~\eqref{impedance} at low driving frequencies. For the extreme case of $\omega = 0$, we have constant driving. Systems with constant driving are already very well-known~\cite{schnakenberg1976}, so any behavior of the currents in the low-frequency limit should coincide with known results. To illustrate, we write the master equation Eq~\ref{master-eqn} in terms of the currents. For the component $P_n(t)$ this reads
\begin{eqnarray}
    \frac{d}{dt}P_n(t)&=&-\sum_{m \neq n}W_{mn}P_n(t)+\sum_{m \neq n}W_{nm}P_m(t)\nonumber \\
    &=& \sum_{m \neq n}J_{nm}(t) \;.
\end{eqnarray}
The term on the RHS sums up all incoming probability currents towards state $n$. When $\omega =0$, the system reaches a time-independent stationary state with $\dot{P_n}(t)=0 \;\; \forall n$, implying Kirchhoff's current law~\cite{schnakenberg1976}
\begin{equation}
    \sum_{m \neq n}J_{nm}=0.
\end{equation}
For small but nonzero frequencies, we leverage the analogous circuit in Figure~\ref{fig:setup_circuit}, dropping the indices $mn$ for notational simplicity. For each branch in the circuit, the conductance is given by
\begin{equation}
    \sigma_{k} =
    \begin{cases}
    R_1^{-1} \qquad & k = 1, \\
    i \omega C_k{(1 + i \omega R_k C_k)}^{-1} \qquad & k \geq 2 .\end{cases}
\end{equation}
In the low frequency limit $\omega \ll {(R_k C_k)}^{-1}$, this becomes
\begin{equation}
    \sigma_{k} =
    \begin{cases}
    R_1^{-1} = A^{(1)}\qquad & k = 1, \\
    i \omega C_k = - i \omega \frac{A^{(k)}}{\lambda_k} \qquad & k \geq 2 .\end{cases}
\end{equation}

Since the conductivity is additive over parallel branches, this gives the following approximation in the low frequency limit:
\begin{equation}
\sigma_{mn} (\omega) \approx A^{(1)}_{mn} - i \omega \sum \limits_{k = 2}^N \frac{A^{(k)}_{mn}}{\lambda_k}.
\end{equation}
Two features become apparent from this expression: firstly, for $\omega = 0$, only the $R_1^{-1} = A^{(1)}$ term contributes, and the system is purely resistive. 
Secondly, when $\omega$ increases, the real part of the conductivity remains approximately constant as the frequency is varied. The imaginary part can increase or decrease from zero, depending on the system and the specific transition. 

%%%% SECTION 5 %%%%%%%%%%%%%%%%%%%%%%%%
\section{The high frequency limit}\label{sec:high-frequency}
In the limiting case $\omega \to \infty$, the conductivity of each branch $k \geq 2$ becomes
\begin{equation}
    \lim_{\omega \to \infty} \sigma_k =  \lim_{\omega \to \infty} {\left(R_k + \frac{1}{i\omega C_k}\right)}^{-1} = \frac{1}{R_k}.
\end{equation}
Borrowing conventional wisdom from electrical circuits, we interpret these results as the fact that the capacitors become perfect conductors as the driving frequency tends to infinity. The disappearance of the capacitors in the equivalent circuit as $\omega \to \infty$ ensures that the imaginary part of the impedance will approach 0, as the capacitors give the only complex contributions. The resulting total conductivity is then
\begin{equation} \label{eq:sigma_infty}
    \sigma_{mn}(\infty) = \lim \limits_{\omega \rightarrow \infty} \sigma_{mn}(\omega) = \sum_{k = 1}^N \frac{1}{R_{mn,k}} = \sum_{k = 1}^N A^{(k)}_{mn}.
\end{equation}
A more careful approximation in the high frequency limit $\omega \gg {(R_k C_k)}^{-1}$ gives
\begin{eqnarray}
     \sigma_k &=& \frac{1}{R_k} {\left(1 + \frac{1}{i\omega R_k C_k}\right)}^{-1} \nonumber \\
     &\approx& \frac{1}{R_k} - \frac{1}{i \omega C_k R_k^2} = A^{(k)} + \frac{A^{(k)} \lambda_k}{i \omega},
\end{eqnarray}
yielding a total conductivity of
\begin{equation}
    \sigma_{mn}(\omega) \approx \sigma_{mn}(\infty) + \frac{1}{i \omega} \sum \limits_{k = 2}^{\infty} A^{(k)}_{mn} \lambda_k.
\end{equation}
As in the low frequency limit, we find that the circuit is, by approximation, purely resistive: as the driving frequency tends to infinity, the imaginary part of the conductivity tends to 0. Contrary to the low frequency limit, we do not find that the real part needs to remain approximately constant as this imaginary part approaches zero. In the present context, it is natural to try to find an expression for $\sigma_{mn}(\infty)$ in terms of the transition rates of the system. To this end, we need to examine the expressions for the coefficients $A^{(k)}_{mn}$. It turns out we can write
\begin{equation}
    A^{(1)}_{mn} = \altbar{W}_{mn} P_n^{\text{eq}} - \altbar{W}_{nm}P_m^{\text{eq}} - \sum_{k=2}^N A^{(k)}_{mn},
\end{equation}
and so
\begin{equation}
    \sigma_{mn}(\infty) = \altbar{W}_{mn} P_n^{\text{eq}} - \altbar{W}_{nm}P_m^{\text{eq}}.
\end{equation}
This expression reveals an interesting relation between $\sigma_{mn}(\infty)$ and the driving $F g(t)$: for any undriven transition, $\altbar{W}_{mn} = \altbar{W}_{nm} = 0$, and so $\sigma_{mn}(\infty) = 0$. In other words, as $\omega \to \infty$, the impedance diverges for undriven transitions and $J_{mn} = 0$. It follows that if the natural transition rates of the system are very slow compared to the driving frequency, the system cannot keep up, and any transition not actively driven will not have an associated current.
The same expression also allows us to rewrite
\begin{equation}
    \sigma_{mn}(\omega) = \sigma_{mn}(\infty) + \sum \limits_{k = 2}^N \frac{A^{(k)}_{mn} \lambda_k}{i \omega - \lambda_k}, \label{eq:high-frequency-exact}
\end{equation}
which allows for easier computation of $\sigma_{mn}(\omega)$. A detailed derivation of the above equations is provided in the Appendix~\ref{app:calc}.

\section{Composite currents}\label{sec:composite}

The impedance~\eqref{impedance} is an expression for the probability current associated with a transition from state $n$ to $m$. In certain systems, such a transition can be the result of several (microscopic) events. For example, if the system consists of physical sites that can hold multiple particles, then there will be multiple transitions corresponding to the same physical displacement of particles. The physical current between two states will be the sum of multiple currents of the form $J_{mn}$ in 
~\eqref{general-current}.

Additionally, transitions within the system can occur through various mechanisms. Think of a transition that can be induced by different thermal baths. Suppose one is interested in the current associated with only a single mechanism, the expression~\eqref{general-current} no longer applies, since that inherently sums all possible mechanisms that make up the transition $n \rightarrow m$. To allow the distinction of different transition mechanisms, we can split the transition matrix $\mathbf{W}$ into different terms,
\begin{equation}
\mathbf{W} = \sum \limits_l \mathbf{W}^{(l)},
\end{equation}
where $\mathbf{W}^{(l)}$ is the transition matrix containing the transition rates associated with the mechanism $l$. Similarly, we 
can write
\begin{align}
\mathbf{W}^{\text{eq}} = \sum \limits_l \mathbf{W}^{\text{eq}, (l)}, \quad
\altbar{\mathbf{W}} = \sum \limits_l \altbar{\mathbf{W}}^{(l)},
\end{align}
for the same decomposition of the matrices $\mathbf{W}^{\text{eq}}$ and $\altbar{\mathbf{W}}$. We can now define a current associated with only a single mechanism as
\begin{equation}\label{current-def-general}
J^{(l)}_{mn} (t) = W^{(l)}_{mn} (t) P_n (t) - W^{(l)}_{nm} (t) P_m (t),
\end{equation}
such that we can recover our previous definition of current~\eqref{general-current} as
\begin{equation}
J_{mn} = \sum \limits_l J^{(l)}_{mn}.
\end{equation}
If we now repeat the second part of Section~\ref{sec:model} for the new current $J^{(l)}_{mn}$, we will find that we can write an 
expansion for its impedance similar to~\eqref{impedance}, but the coefficients $A^k_{mn}$ will gain an additional index to signify that they belong to a single transition mechanism,
\begin{eqnarray}\label{coeffs-general}
&&A_{mn}^{(1, l)} = W_{mn}^{\text{eq}, (l)} \altbar{P}_n^{\text{ad}} + \altbar{W}_{mn}^{(l)} P_n^{\text{eq}} \nonumber \\
&& \hspace{3cm}- W_{nm}^{\text{eq}, (l)} \altbar{P}_m^{\text{ad}} - \altbar{W}_{nm}^{(l)} P_m^{\text{eq}}\; ; \\
&&A_{mn}^{(k \geq 2, l)} =-\inner{\mathbf{v}_k}{\altbar{\mathbf{P}}^{\text{ad}}} \left(W_{mn}^{\text{eq}, (l)} v_{k, n} - W_{nm}^{\text{eq}, (l)} v_{k, m} \right). \nonumber
\end{eqnarray}
From these considerations, we see that any desired sum of the currents $J_{mn}^{(l)}$ can be found by finding the appropriate coefficients $A_{mn}^{(k, l)}$ and summing all coefficients with the same value for $k$. It is then the result of these sums that will appear in the place of the $A^{(k)}_{mn}$ in~\eqref{impedance} when calculating a composite current. 
%%%%%%%%%%%%%%%%%%%%%%%%%%%%%%%%%%%%%%%%%%%%%%%%%%%%%%%%%%
\section{Illustration of the method}\label{sec:example}
%%%%%%%%%%%%%%%%%%%%%%%%%%%%%%%%%%%%%%%%%%%%%%%%%%%%%%%%%%
\subsection{Linear hopping chain}
\begin{figure}
    \centering
    \includegraphics[width=0.95\columnwidth]{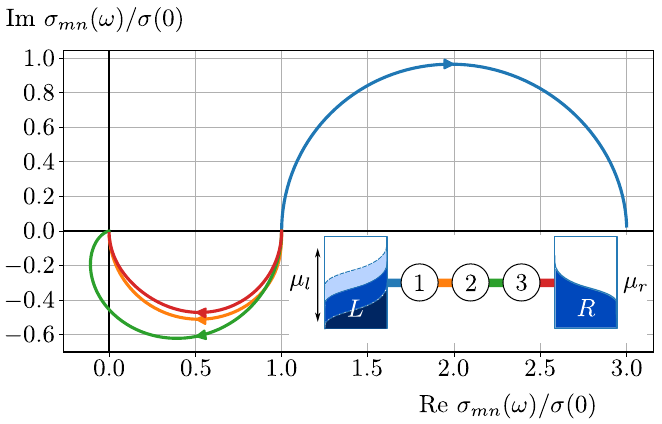}
    \caption{The conductivities $\sigma_{mn}$ for all the possible transitions in a linear chain of three sites as a function of the driving frequency $\omega$, normalized by the adiabatic conductivity $\sigma(0)$ (the subscripts have been dropped as $\sigma(0)$ is the same for all transitions). The driving acts only on the transition to the left reservoir. The arrows on the curves indicate the direction of increasing frequency. The transitions shown in the inset are colored according to their conductivity curves. Note that the subscripts of $\sigma(0)$ have been dropped as they are the same for all transitions.}\label{fig:linear-chain-conds}
\end{figure}

As an illustration of the previous results, consider a linear chain of $N$ hopping sites, with each end of the chain connected to a 
particle reservoir. The inset in Figure~\ref{fig:linear-chain-conds} shows an example for $N = 3$. A particle can enter and exit the chain from either reservoir and can hop between neighboring sites. If the chemical potentials of both reservoirs are equal, the system is in thermal equilibrium. Varying either or both chemical potentials will drive the system out of equilibrium. For simplicity, let us consider a single-particle hopping model on this chain, i.e.\ there can only be one particle in the chain at any given time. The resulting system then has $N+1$ possible states if the chain has $N$ sites. For further simplification, we will take the hopping rates between adjacent sites to be unity, and the hopping rates between the reservoirs and the chain to take on a  Fermi distribution form,
\begin{equation}\label{eq:fermi-rates}
    \Gamma_+ = \frac{1}{1 + e^{- \mu}}, \qquad \Gamma_- = 1 - \Gamma_+.
\end{equation}
Here, $\Gamma_+$ and $\Gamma_-$ are the rates for the particle to jump into or out of the system, respectively, and $\mu$ is the chemical 
potential of the reservoir. We have set the inverse temperature $\beta = 1$ and will keep this convention throughout this section. 
Furthermore, the chemical potential of the right reservoir will be set to zero, while that of the left reservoir will be set to oscillate 
periodically around zero. This makes the transition between the left reservoir and the chain the only driven transition in the system.

Using the framework outlined above, one can then calculate the linearized current at each of the transitions between adjacent sites, as well as the currents between the chain and the reservoirs, all as a function of the driving frequency. Plotting all the associated conductivities in the complex plane as a curve parametrized by the frequency, yields graphs like Figure~\ref{fig:linear-chain-conds}.

\begin{figure}[t]
    \centering
    \includegraphics[width=0.95\columnwidth]{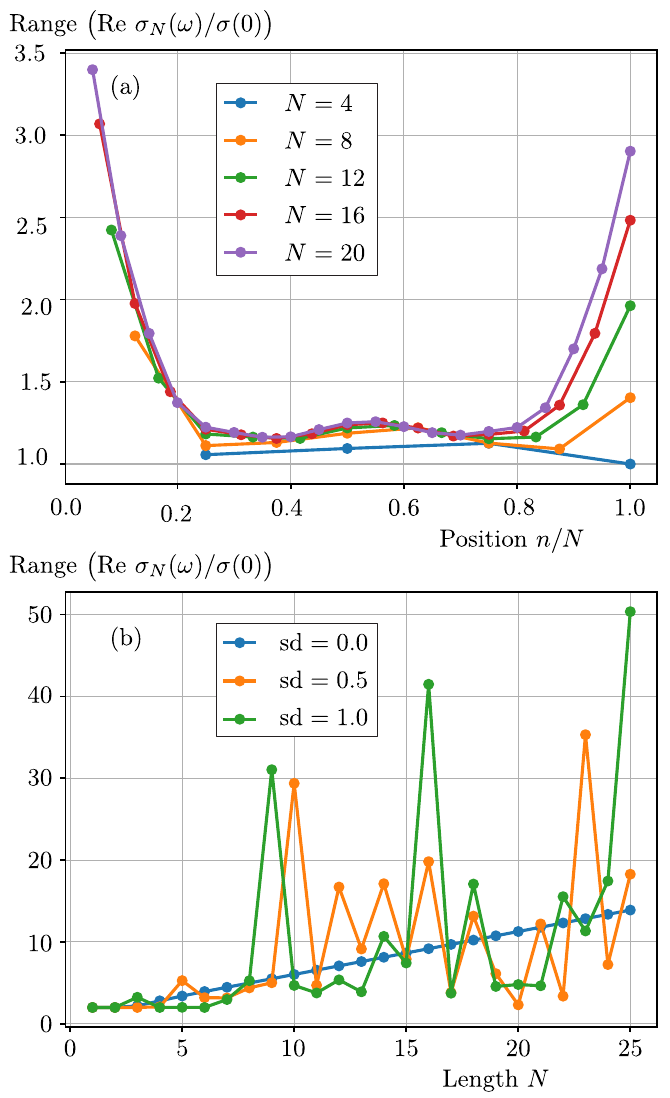}
    \caption{(a) The size of the conductivity plot as a function of relative position $n/N$ along the chain for different chain lengths $N$. The conductivity for the transition $n \rightarrow n+1$ (or $N \rightarrow R$ if $n = N$) is denoted by $\sigma_n$. The conductivity $\sigma_0$ for the transition with the left reservoir isn't shown because, being a driven transition, its size is more determined by its high-frequency limit than by its shape.\\
    (b) The size of the conductivity plot for the current between the system and the undriven reservoir, as a function of the chain length $N$, for an ideal system and two disordered systems in which the energy was sampled from normal distributions with standard deviations (sd) of $0.5$ and $1$.}\label{fig:linear-chain-range}
\end{figure}

We see, as already calculated in section~\ref{sec:low-frequency}, that for $\omega$ going to zero, $\mbox{Im}(\sigma)$ also goes to zero. Furthermore, for all curves, the real part of the conductivity is approximately constant for small $\omega$. In accordance with Section~\ref{sec:high-frequency}, we see that the conductivities of the non-driven transitions tend to zero in the high-frequency limit. As seen for the $2 \to 3$ transition, it need not be that the curve approaches the real axis `vertically' in the high frequency limit.
Additionally, these conductivity graphs depend on the position along the chain. For example, the ranges of the real and imaginary parts of the conductivity are different for most of the curves. As an illustration, we calculate the range of the real part for all transitions. We do this for different chain sizes and compare the results by dividing by the adiabatic conductivity $\sigma (0)$, which is the same for all currents in a given system, but not necessarily for currents in different systems. The results of this comparison are shown in Figure~\ref{fig:linear-chain-range}.\\

There seems to be a relation between the length of the chain and the size of the conductivity plot associated with the transition connected to the right reservoir. Calculating this specific conductivity for different sizes reveals a linear behavior for systems with more than $\sim 5$ sites, which is already suggested by plot (a) of Figure~\ref{fig:linear-chain-range}. The result is shown in plot (b) of the same Figure.\\

These results were calculated by numerically diagonalizing the matrices $\mathbf{W}^{\text{eq}}$ associated with those systems. All systems were assumed to be ideal, with transitions between sites occurring at a constant rate of unity and transitions to and from reservoirs following a Fermi distribution rate~\eqref{eq:fermi-rates}. However, it would be interesting to see if this behavior survives in systems without this idealized structure. It is possible to generate non-ideal systems by associating energies with each site and modifying the transition rates so that they still obey detailed balance with respect to the Boltzmann distribution for those energies. Specifically, we choose the rates between sites to be
\begin{equation}
W^{	ext{eq}}_{mn} = e^{-\frac{E_m - E_n}{2}},
\end{equation}
with $E_n$ the energy of site $n$. The rates~\eqref{eq:fermi-rates} involving reservoirs are similarly modified, now involving the difference between the chemical potential and the energy of the relevant site.

With these rates, we can generate a random array of energies and repeat the calculations for chains of various lengths, where a chain of length $N$ has sites whose energies are equal to the first $N$ randomly generated energies. The result of this is drawn in Figure ~\ref{fig:linear-chain-range}. The energies used were drawn from a normal distribution centered around zero with various standard deviations. We see that the linear behavior still largely holds, but becomes increasingly obscured as the standard deviation (and thus the degree of randomness) increases.

\subsection{Linear periodic chain with shortcuts}
Equations~\eqref{impedance} and~\eqref{coeffs-simple} have an explicit dependence on the microscopic details of the system through the appearance of all eigenvalues and eigenvectors. The question of whether an impedance measurement could distinguish between two similar systems naturally arises. As it turns out, in some cases it can. As an illustration, we consider a simple linear chain again, but this time we apply periodic boundary conditions. The perturbation will be given by a uniform force field $F$, which modifies the 
transition rates by
\begin{align}\label{electric-field-rates}
W_{n, n+1} (F) &= W_{n, n+1} (F = 0) ~ e^{-F}, \\
W_{n+1, n} (F) &= W_{n+1, n} (F = 0) ~ e^{F},
\end{align}
where we omitted the periodic time dependence of $F$ to ease notation. The periodic boundary conditions identify site $N+1$ with site $1$. 
At $F = 0$, the matrix $\mathbf{W}^{\text{eq}} = \mathbf{W}(F = 0)$ is assumed to satisfy detailed balance.

\begin{figure}[t]
\centering
\includegraphics[width=0.95\columnwidth]{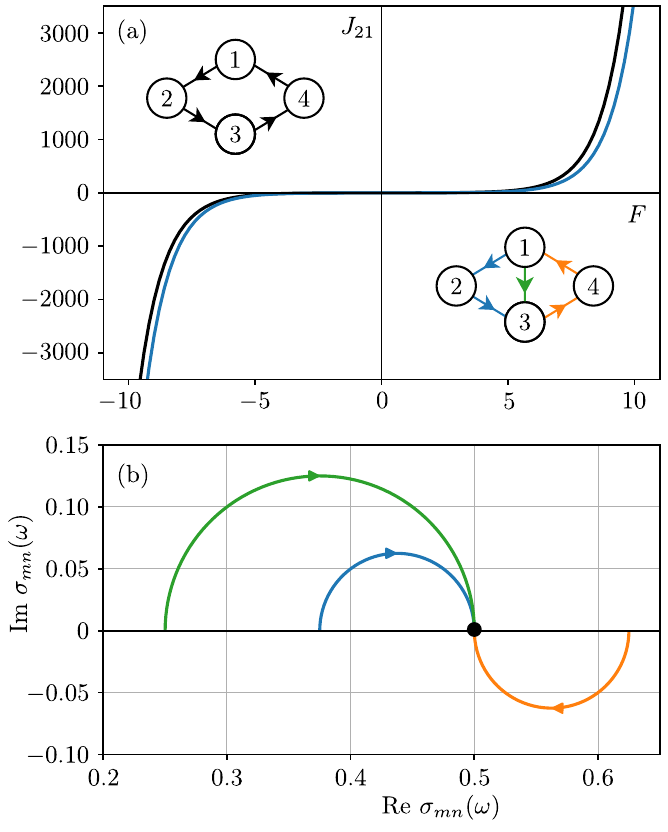}
\caption{(a) The stationary current $J_{21}$ as a function of the applied constant driving $F$ outside the linear regime for both the system with and without a shortcut. (b) The conductivities for the periodic chain with a shortcut (indicated by the arrow between states 1 and 3 in the inset on (a)). Arrows in the inset indicate the direction of the driving at zero frequency, where the current is considered positive. The arrows on the curves indicate the direction of increasing frequency. Currents with the same color on the inset are identical by symmetry. The dot represents the system's conductivity without a shortcut.}\label{fig:linear-chain-comparison}
\end{figure}

Specifically, we will consider a chain of $N=4$ sites, with all undriven transition rates set to unity. We compare it to the same system with a shortcut between sites $1$ and $3$. This transition is modified similarly to the others,
\begin{align}
W_{1, 3} (F) &= W_{1, 3} (F = 0) ~ e^{-F}\\
W_{3, 1} (F) &= W_{3, 1} (F = 0) ~ e^{F}.
\end{align}
Its undriven transition rate is also taken to be unity in both directions. A graphical representation of this can be seen in the inset in Figure~\ref{fig:linear-chain-comparison}.

If one were able to measure the current through different transitions, the violation of Kirchhoff's current law would immediately reveal the presence of this shortcut. However, if not all currents are observable, this may not be an option. As illustrated below, the conductivities of these systems differ markedly. This difference allows us to distinguish the two systems based on a single measurement of stochastic 
impedance.

From the transition rates defined in~\eqref{electric-field-rates}, the matrix $\altbar{\mathbf{W}}$ can be found, since 
\begin{equation}\label{electric-field-perturbation}
\altbar{W}_{n, n+1} = \frac{\partial W_{n, n+1}(F) }{\partial F} \bigg|_{F = 0} = - W^{\text{eq}}_{n, n+1} = - \altbar{W}_{n+1, n}
\end{equation}
for all $n$ in the system without a shortcut. In the system with a shortcut, the additional transition causes two more matrix elements in $\altbar{\mathbf{W}}$ to be nonzero. Those can be found similarly. This structure makes the perturbation an abstraction of the effect that a uniform electric field has on the hopping of a charged particle~\cite{ishii1985}.

Note, however, that the above antisymmetry only holds for the off-diagonal elements of the $\altbar{\mathbf{W}}$. In a slightly more generalized notation, we can write them as $\altbar{W}_{nm} = a_{nm} W^{	ext{eq}}_{nm}$ for $m \neq n$, where the $a_{mn}$ are antisymmetric. The matrix $\altbar{\mathbf{W}}$ is only antisymmetric if its diagonal elements are zero. This is the case in the periodic chain without a shortcut, since the off-diagonal elements in every column will have exactly two nonzero entries, which are opposite to each other. In general, however, this will not be the case.
We can now calculate the matrix product $\altbar{\mathbf{W}} \mathbf{P}^{\text{eq}}$ appearing in~\eqref{coeffs-simple} as

\begin{align}
{[\altbar{\mathbf{W}} \mathbf{P}^{\text{eq}}]}_n &= \sum \limits_{m = 1}^4 \altbar{W}_{nm} P^{\text{eq}}_{m}\\
&= \altbar{W}_{nn} P^{\text{eq}}_{n} + \sum \limits_{m \neq n} a_{nm} W^{\text{eq}}_{nm} P^{\text{eq}}_{m}\\
&= \altbar{W}_{nn} P^{\text{eq}}_{n} - \sum \limits_{m \neq n} a_{mn} W^{\text{eq}}_{mn} P^{\text{eq}}_{n}\\
&= \altbar{W}_{nn} P^{\text{eq}}_{n} -  \sum \limits_{m \neq n} \altbar{W}_{mn} P^{\text{eq}}_{n}= 2 P^{\text{eq}}_{n} \altbar{W}_{nn}
\end{align}
where we used the antisymmetry property of the $a_{mn}$ and the fact that the columns of $\altbar{\mathbf{W}}$ must still sum to zero. For the system 
without shortcut, we have $\altbar{W}_{nn} = 0$ for all $n$, so $\altbar{\mathbf{W}} \mathbf{P}^{\text{eq}} = 0$. It immediately follows that $A^k_{mn} = 0$ for all $k \neq 1$. 
Hence, the current is independent of frequency, and the conductivities become those of a simple resistor if there is no shortcut.

In the system with a shortcut, however, the additional driven transition breaks the antisymmetry of $\altbar{\mathbf{W}}$. In that case, the coefficients~\eqref{coeffs-simple} will generally not be zero and the current gains a dependence on the driving frequency. Such a dependence gives rise to a qualitative difference between the conductivities of the two systems. To illustrate that the system with a shortcut does depend on frequency, Figure~\ref{fig:linear-chain-comparison} shows the associated conductivities for all its transitions.

This difference is barely measurable using a constant force. One can calculate the stationary current for constant driving outside of the linear regime by finding the null space of $\mathbf{W} (F)$ as a function of (the amplitude of) $F$. Because $\mathbf{W} (F) \in \mathbb{R}^{4\times 4}$, this is analytically possible. The resulting expressions are not very insightful, but a comparison of the two graphs, which can be seen in figure~\ref{fig:linear-chain-comparison}, shows there is barely any difference between the two systems in this regard.

It is important to note that the shortcut considered here is not the only possible modification that could cause the antisymmetry of $\altbar{\mathbf{W}}$ to break. It is not hard to see that a similar breaking of the antisymmetry happens when the equilibrium transition rates are not all equal. This can be accomplished, for example, by assigning different energies to each site and adjusting the transition rates to satisfy detailed balance. In this regard, the above discussion is primarily a showcase of the potential to detect such broken 
symmetry.
%%%%%%%%%%%%%%%%%%%%%%%%%%%%%%%%%%%%%%%%%%%
\section{Conclusion}\label{sec:conclusion}
We have derived an expression for probability currents in general stochastic systems. We defined stochastic impedance in an analogous manner to that of electric circuits. We have investigated the low- and high-frequency limits of the results, and generalized them to include more complicated current 
expressions. We finished with two illustrative applications of the framework to simple hopping systems, showing interesting behavior and 
outlining its potential to reveal broken symmetries.
%%%%%%%%%%%%%%%%%%%%%%%%%%%%%%%%%%%%%%%%%%%
%%%%%%%%%%%%%%%%%%%%%%%%%%%%%%%%%%%%%%%%%%%
\appendix
%%%%%%%%%%%%%%%%%%%%%%%%%%%%%%%%%%%%%%%%%%%
%%%%%%%%%%%%%%%%%%%%%%%%%%%%%%%%%%%%%%%%%%%
\section{Diagonalization of $\mathbf{W}^{\text{eq}}$}\label{app:Diag}
%%%%%%%%%%%%%%%%%%%%%%%%%%%%%%%%%%%%%%%%%%%
We start with a presentation of the details of the diagonalization of $\mathbf{W}^{\text{eq}}$ and subsequently, the calculation of the matrix product $e^{\mathbf{W}^{\text{eq}} \tau} \altbar{\mathbf{P}}^{\text{ad}}$. We already know that the matrix $\mathbf{W}^{\text{eq}}$ is diagonalizable, so it has a full set of eigenvalues $\lambda_k$ and eigenvectors $\mathbf{v}_k$. Its largest eigenvalue is $\lambda_1 = 0$ and the associated eigenvector is $\mathbf{v}_1 = \mathbf{P}^{\text{eq}}$. However, $\mathbf{W}^{\text{eq}}$ is not self-adjoint with respect to the standard inner product. It is conventional to introduce a new inner product,
\begin{equation}\label{inner-app}
\inner{\mathbf{v}}{\mathbf{w}} = \sum \limits_{n = 1}^{N} \frac{v_n w_n}{P^{\text{eq}}_n},
\end{equation}
with respect to which the matrix $\mathbf{W}^{\text{eq}}$ is self adjoint. We can then choose the eigenvectors to be orthonormal with respect to this inner product.

The matrix $\mathbf{W}^{\text{eq}}$ can now be diagonalized as $\mathbf{W}^{\text{eq}} = \mathbf{S} \mathbf{D} \mathbf{S}^{-1}$, where $\mathbf{D}$ is a diagonal matrix whose elements are $D_{ii} = \lambda_i$ and 
\begin{equation}
S = \begin{pmatrix} \mathbf{v}_1 & \dots & \mathbf{v}_N \end{pmatrix} \; , \qquad S^{-1} = \begin{pmatrix} \mathbf{w}_1 \\ \vdots \\ \mathbf{w}_N \end{pmatrix}.
\end{equation}
The vectors $\mathbf{w}_k$ are the left eigenvectors of $\mathbf{W}^{\text{eq}}$. They are related to the right eigenvectors by
\begin{equation}
w_{k, n} = \frac{v_{k, n}}{P^{\text{eq}}_n},
\end{equation}
where $w_{k, n}$ is the $n-$th component of $\mathbf{w}_k$, and similarly for $v_{k, n}$. Using this relation along with the inner product~\eqref{inner-app} and the requirement that the eigenvectors are orthonormal with respect to it, we see that $\mathbf{S}^{-1}$ is indeed the inverse of $\mathbf{S}$. The matrix exponential can now be calculated as $e^{\mathbf{W}^{\text{eq}} \tau} = e^{\mathbf{S} \mathbf{D} \mathbf{S}^{-1} \tau} = \mathbf{S} e^{\mathbf{D} \tau} \mathbf{S}^{-1}$. Carrying out the matrix multiplications, we get
\begin{align}
\left[ \mathbf{S}^{-1} \altbar{\mathbf{P}}^{\text{ad}} \right]_n &= \inner{\mathbf{v}_n}{\altbar{\mathbf{P}}^{\text{ad}}}, \\
\left[ \mathbf{S} e^{\mathbf{D}\tau} \right]_{mn} &= v_{n, m} e^{\lambda_n \tau},
\end{align}
where $v_{m, n}$ is the $n-$th component of the eigenvector $\mathbf{v}_m$. It follows that the matrix exponential is given by
\begin{equation}
e^{\mathbf{W}^{\text{eq}} \tau} \altbar{\mathbf{P}}^{\text{ad}} = \sum \limits_{n = 2}^N \mathbf{v}_n e^{\lambda_n \tau} \inner{\mathbf{v}_n}{\altbar{\mathbf{P}}^{\text{ad}}},
\end{equation}
where we can start the sum at $n = 2$ because $\mathbf{v}_1 = \mathbf{P}^{\text{eq}}$, so we must have $\inner{\mathbf{v}_1}{\altbar{\mathbf{P}}^{\text{ad}}} = 0$ by orthogonality.

\section{Degeneracies}\label{app:degens}

The coefficients $A^{(k)}_{mn}$ used to calculate the conductivities are not manifestly independent of the choice of eigenbasis for $\mathbf{W}^{\text{eq}}$, even after requiring the basis to be orthonormal. Especially when one or more eigenvalues of $\mathbf{W}^{\text{eq}}$ are degenerate, the resulting freedom in the choice of eigenbasis should not affect conductivity, since that is a physical quantity.

To prove that the conductivities are independent of the choice of orthonormal eigenbasis of $\mathbf{W}^{\text{eq}}$, it is sufficient to show this is true for a single eigenspace; the full space $\mathbb{R}^N$ is simply the union of all the eigenbases for the eigenspaces of $\mathbf{W}^{\text{eq}}$. In the following, we prove that a change of basis in the eigenspace of a possibly degenerate eigenvalue does not change the conductivity. Since we are focusing on a single current, we will now drop the subscripts $mn$ for notational simplicity whenever they are not needed.

Take any eigenvalue $\lambda_k$ for some $k$. Call its associated eigenspace $E_k$. $\lambda_k$ may or may not be degenerate, so $E_k$ has some undetermined dimension $d \leq N-1$. Then, in order to calculate the coefficients $A^{(k)}$, we must choose an orthonormal basis $\beta$ for $E_k$ and use those basis vectors to calculate the $d$ associated coefficients $A^{(k)}_{\beta}$ to $A^{(k + d-1)}_{\beta}$. The index $\beta$ indicates that these coefficients will depend on the choice of basis $\beta$. Note that whenever we talk about orthogonality in this section, we mean orthogonality with respect to the inner product~\eqref{inner-app}. Because each of the basis vectors in $\beta$ is associated with the same eigenvalue, their associated coefficients can be combined into an effective coefficient $\tilde{A}^{(k)}_{\beta}$, given by
\begin{equation}
\tilde{A}^{(k)}_{\beta} = \sum \limits_{j = 0}^{d-1} A^{(k+j)}_{\beta}.
\end{equation}
Our goal is to show that this effective coefficient is independent of $\beta$. To accomplish this, we will show that either it is zero regardless of $\beta$, or there exists a special basis $\beta'$ for which $\tilde{A}^{(k)}_{\beta} = \tilde{A}^{(k)}_{\beta'}$ for any $\beta$.

First, consider the subspace $V^{\perp} = \{\mathbf{v} \, | \, \inner{\mathbf{v}}{\altbar{\mathbf{P}}^{\text{ad}}} = 0\}$ of $\mathbb{R}^{N}$, which consists of all vectors perpendicular to $\altbar{\mathbf{P}}^{\text{ad}}$. This space has dimension $N-1$ as it is orthogonal to a one-dimensional subspace. We can construct the intersection $E_k \cap V^{\perp}$, whose dimension we can calculate as
\begin{align}
\dim (E_k \cap V^{\perp} ) &= \dim E_k + \dim V^{\perp} - \dim (E_k + V^{\perp}) \\
&\geq d-1,
\end{align}
since $\dim E_k = d$ and $E_k + V^{\perp}$ is either all of $\mathbb{R}^{N}$ or just $V^{\perp}$. Obviously we also have $\dim E_k \cap V^{\perp} \leq \dim E_k = d$. Hence, the dimension of $E_k \cap V^{\perp}$ can only be either $d$ or $d-1$. In the first case $E_k \subset V^{\perp}$ and so $\inner{\mathbf{v}}{\altbar{\mathbf{P}}^{\text{ad}}} = 0$ for all $\mathbf{v} \in E_k$ and therefore $\tilde{A}^{(k)}_{\beta} = 0$ for any choice of $\beta$.\\
Next, consider the case where $\dim E_k \cap V^{\perp} = d-1$. Then we can construct an orthonormal basis $\tilde{\beta} = \{\mathbf{v}_k \, | \, k = 1, \dots d-1 \}$ for this intersection. There exists a vector $\mathbf{v}_d \in E_k \setminus V^{\perp}$ such that the basis $\beta' = \tilde{\beta} ~ \cup ~ \{\mathbf{v}_d \}$ is an orthonormal basis for $E_k$. Because $E_k \setminus V^{\perp}$ is a one-dimensional space, and $\mathbf{v}_k$ is a normal vector, it is unique up to a sign difference.\\

By construction, $\inner{\mathbf{v}_k}{\altbar{\mathbf{P}}^{\text{ad}}} = 0$ for all $k \neq d$. Hence, we have
\begin{align}
\tilde{A}^{(k)}_{\beta'} &= -\inner{\mathbf{v}_d}{\altbar{\mathbf{P}}^{\text{ad}}} ( W_{mn}^{\text{eq}} v_{d, n} - W_{nm}^{\text{eq}} v_{d, m}) \notag \\ &= -\inner{\mathbf{v}_d }{\altbar{\mathbf{P}}^{\text{ad}}} I(\mathbf{v}_d) ,
\end{align}
where we have contracted the second factor into a linear function,
\begin{equation}\label{function-i-app}
I : \mathbb{R}^{N} \rightarrow \mathbb{R}: \mathbf{v} \mapsto W^{\text{eq}}_{mn} v_{n} - W^{\text{eq}}_{nm} v_{m}.
\end{equation}
This will shorten the notation. Note that the only choices to make for $\beta'$ are the vectors in $\tilde{\beta}$ and the sign of $\mathbf{v}_d$. These choices are independent of each other, and the effective coefficient doesn't depend on either.

It remains to show that the effective coefficient calculated in any other basis is equal to this one. To this end, consider any orthonormal basis $\beta = \{ \mathbf{w}_k ~ | ~ k = 1 , \dots, d \}$ for $E_k$. We can convert between $\beta$ and $\beta'$ by
\begin{equation}
\mathbf{w}_k = \sum \limits_{l = 1}^d c_{kl} \mathbf{v}_l = \sum \limits_{l = 1}^d \inner{\mathbf{w}_k}{\mathbf{v}_l} \mathbf{v}_l.
\end{equation}
We can now calculate, for any such basis $\beta$,
\begin{align*}
\tilde{A}^{(k)}_{\beta} &= \sum \limits_{l = 0}^{m-1} A^{(k+l)}_{\beta}= - \sum \limits_{l = 1}^d \inner{\mathbf{w}_l}{\altbar{\mathbf{P}}^{\text{ad}}} I(\mathbf{w}_l)\\
&= - \sum \limits_{l = 1}^d \inner{c_{ld} \mathbf{v}_d}{\altbar{\mathbf{P}}^{\text{ad}}} I (\mathbf{w}_l)\\
&= - \sum \limits_{l = 1}^d \inner{\mathbf{v}_d}{\altbar{\mathbf{P}}^{\text{ad}}} I (\inner{\mathbf{w}_l}{\mathbf{v}_d} \mathbf{w}_l)\\
&= - \inner{\mathbf{v}_d}{\altbar{\mathbf{P}}^{\text{ad}}} I (\mathbf{v}_d)= \tilde{A}^{(k)}_{\beta'}.
\end{align*}
This result concludes the proof. It follows that the current is independent of the choice of basis for eigenspaces of degenerate eigenvalues.\qed
%%%%%%%%%%%%%%%%%%%%%%%%%%%%%%%%%%%%%%%%%%%
\section{Calculation of the conductivities}\label{app:calc}
%%%%%%%%%%%%%%%%%%%%%%%%%%%%%%%%%%%%%%%%%%%
To rewrite the coefficient $A^{(1)}_{mn}$, we use the expansion of $\altbar{\mathbf{P}}^{\text{ad}}$ in terms of the eigenvectors of $\mathbf{W}^{\text{eq}}$:
\begin{equation}
    \altbar{\mathbf{P}}^{\text{ad}} = \sum_{k=2}^N \inner{\mathbf{v}_k}{\altbar{\mathbf{P}}^{\text{ad}}} \mathbf{v}_k.
\end{equation}
Note that the $k=1$ term is not included, because $\inner{\mathbf{v}_k}{\altbar{\mathbf{P}}^{\text{ad}}} = 0$. Substituting this into parts of $A^{(1)}_{mn}$, we find
\begin{equation}
    W^{\text{eq}}_{mn} \altbar{P}^{\text{ad}}_{n} - W^{\text{eq}}_{nm} \altbar{P}^{\text{ad}}_{m} = - \sum_{k=2}^N A^{(k)}_{mn}
\end{equation}
and so
\begin{equation}
    A^{(1)}_{mn} = \altbar{W}_{mn} P^{\text{eq}}_{n} - \altbar{W}_{nm} P^{\text{eq}}_{m} - \sum_{k=2}^N A^{(k)}_{mn}
\end{equation}
From this point, it is elementary to obtain the expressions for $\sigma_{mn}(\infty)$ and $\sigma_{mn}(\omega)$:
\begin{align}
    & \sigma_{mn}(\infty) = \sum_{k=1}^N A^{(k)}_{mn} = \altbar{W}_{mn} P^{\text{eq}}_{n} - \altbar{W}_{nm} P^{\text{eq}}_{m}\\
    & \sigma_{mn}(\omega) = \sum_{k=1}^N \frac{A^{(k)}_{mn}i\omega}{i\omega - \lambda_k} = \sigma_{mn}(\infty)+ \sum_{k=2}^N \frac{A^{(k)}_{mn} \lambda_k}{i\omega - \lambda_k}.
\end{align}
The main advantage of these formulae is the elimination of terms containing $\altbar{\mathbf{P}}^{\text{ad}}$, as these are more difficult to calculate than, for example, terms containing $\altbar{\mathbf{W}}$. At this point, we have not completely succeeded in this endeavor. For $k \geq 2$, the expressions for $A^{(k)}_{mn}$ still contain the expansion coefficients $\inner{\mathbf{v}_k}{\altbar{\mathbf{P}}^{\text{ad}}}$. To eliminate these, we start from
\begin{equation}
    0 = \mathbf{W}(t) \mathbf{P}^{\text{ad}}(t)= (\altbar{\mathbf{W}} \mathbf{P}^{\text{eq}} + \mathbf{W}^{\text{eq}} \altbar{\mathbf{P}}^{\text{ad}}) F g(t),
\end{equation}
where the second equality holds assuming linear response. We see that $\mathbf{W}^{\text{eq}} \altbar{\mathbf{P}}^{\text{ad}} = - \altbar{\mathbf{W}} \mathbf{P}^{\text{eq}}$. We can now use the diagonalization of $\mathbf{W}^{\text{eq}}$ given in the first section to obtain:
\begin{equation}
    \mathbf{D} \mathbf{S}^{-1} \altbar{\mathbf{P}}^{\text{ad}} = - \mathbf{S}^{-1} \altbar{\mathbf{W}} \mathbf{P}^{\text{eq}}.
\end{equation}
Analyzing the components of this vector, we find
\begin{align}
    \left[ \mathbf{D} \mathbf{S}^{-1} \altbar{\mathbf{P}}^{\text{ad}} \right]_{k} &= \left[ - \mathbf{S}^{-1} \altbar{\mathbf{W}} \mathbf{P}^{\text{eq}} \right]_{k}\\
    \inner{\mathbf{v}_k}{\altbar{\mathbf{P}}^{\text{ad}}} &= - \frac{\inner{\mathbf{v}_k}{\altbar{\mathbf{W}} \mathbf{P}^{\text{eq}}}}{\lambda_k}.
\end{align}
Note that the second step is not allowed for $k = 1$, since $\lambda_1 = 0$. This does not matter, however, since we already know $\inner{\mathbf{v}_1}{\altbar{\mathbf{P}}^{\text{ad}}} = 0$. This gives an expression for the expansion coefficients that does not explicitly contain $\altbar{\mathbf{P}}^{\text{ad}}$.

%%%%%%%%%%%%%%%%%%%%%%%%%%%%%%%%%%%%%%%%%%%
%%%%%%%%%%%%%%%%%%%%%%%%%%%%%%%%%%%%%%%%%%% 
\section*{Acknowledgments}
\noindent
This study was supported by the Special Research Fund (BOF) of Hasselt University (BW grant no. BOF20OWB22, BM grant no. BOF24KP19, JH and BC grant no. BOF25GP02).
%%%%%%%%%%%%%%%%%%%%%%%%%%%%%%%%%%%%%%%%%%%%
%%%%%%%%%%%%%%%%%%%%%%%%%%%%%%%%%%%%%%%%%%%

\bibliography{bib.bib}

\begin{thebibliography}{18}%
\makeatletter
\providecommand \@ifxundefined [1]{%
 \@ifx{#1\undefined}
}%
\providecommand \@ifnum [1]{%
 \ifnum #1\expandafter \@firstoftwo
 \else \expandafter \@secondoftwo
 \fi
}%
\providecommand \@ifx [1]{%
 \ifx #1\expandafter \@firstoftwo
 \else \expandafter \@secondoftwo
 \fi
}%
\providecommand \natexlab [1]{#1}%
\providecommand \enquote  [1]{``#1''}%
\providecommand \bibnamefont  [1]{#1}%
\providecommand \bibfnamefont [1]{#1}%
\providecommand \citenamefont [1]{#1}%
\providecommand \href@noop [0]{\@secondoftwo}%
\providecommand \href [0]{\begingroup \@sanitize@url \@href}%
\providecommand \@href[1]{\@@startlink{#1}\@@href}%
\providecommand \@@href[1]{\endgroup#1\@@endlink}%
\providecommand \@sanitize@url [0]{\catcode `\\12\catcode `\$12\catcode
  `\&12\catcode `\#12\catcode `\^12\catcode `\_12\catcode `\%12\relax}%
\providecommand \@@startlink[1]{}%
\providecommand \@@endlink[0]{}%
\providecommand \url  [0]{\begingroup\@sanitize@url \@url }%
\providecommand \@url [1]{\endgroup\@href {#1}{\urlprefix }}%
\providecommand \urlprefix  [0]{URL }%
\providecommand \Eprint [0]{\href }%
\providecommand \doibase [0]{https://doi.org/}%
\providecommand \selectlanguage [0]{\@gobble}%
\providecommand \bibinfo  [0]{\@secondoftwo}%
\providecommand \bibfield  [0]{\@secondoftwo}%
\providecommand \translation [1]{[#1]}%
\providecommand \BibitemOpen [0]{}%
\providecommand \bibitemStop [0]{}%
\providecommand \bibitemNoStop [0]{.\EOS\space}%
\providecommand \EOS [0]{\spacefactor3000\relax}%
\providecommand \BibitemShut  [1]{\csname bibitem#1\endcsname}%
\let\auto@bib@innerbib\@empty
%</preamble>
\bibitem [{\citenamefont {Green}(1954)}]{Green1954}%
  \BibitemOpen
  \bibfield  {author} {\bibinfo {author} {\bibfnamefont {M.~S.}\ \bibnamefont
  {Green}},\ }\bibfield  {title} {\bibinfo {title} {Markoff random processes
  and the statistical mechanics of time‐dependent phenomena. ii. irreversible
  processes in fluids},\ }\href@noop {} {\bibfield  {journal} {\bibinfo
  {journal} {J. Chem. Phys.}\ }\textbf {\bibinfo {volume} {22}},\ \bibinfo
  {pages} {398} (\bibinfo {year} {1954})}\BibitemShut {NoStop}%
\bibitem [{\citenamefont {Kubo}(1957)}]{Kubo1957}%
  \BibitemOpen
  \bibfield  {author} {\bibinfo {author} {\bibfnamefont {R.}~\bibnamefont
  {Kubo}},\ }\bibfield  {title} {\bibinfo {title} {Statistical-mechanical
  theory of irreversible processes. i. general theory and simple applications
  to magnetic and conduction problems},\ }\href@noop {} {\bibfield  {journal}
  {\bibinfo  {journal} {J. Phys. Soc. Japan}\ }\textbf {\bibinfo {volume}
  {12}},\ \bibinfo {pages} {570} (\bibinfo {year} {1957})}\BibitemShut
  {NoStop}%
\bibitem [{\citenamefont {Hayashi}\ and\ \citenamefont
  {Sasa}(2006)}]{Hayashi2006}%
  \BibitemOpen
  \bibfield  {author} {\bibinfo {author} {\bibfnamefont {K.}~\bibnamefont
  {Hayashi}}\ and\ \bibinfo {author} {\bibfnamefont {S.-i.}\ \bibnamefont
  {Sasa}},\ }\bibfield  {title} {\bibinfo {title} {Linear response theory in
  stochastic many-body systems revisited},\ }\href@noop {} {\bibfield
  {journal} {\bibinfo  {journal} {Physica A}\ }\textbf {\bibinfo {volume}
  {370}},\ \bibinfo {pages} {407 } (\bibinfo {year} {2006})}\BibitemShut
  {NoStop}%
\bibitem [{\citenamefont {Marconi}\ \emph {et~al.}(2008)\citenamefont
  {Marconi}, \citenamefont {Puglisi}, \citenamefont {Rondoni},\ and\
  \citenamefont {Vulpiani}}]{Marconi2008}%
  \BibitemOpen
  \bibfield  {author} {\bibinfo {author} {\bibfnamefont {U.~M.~B.}\
  \bibnamefont {Marconi}}, \bibinfo {author} {\bibfnamefont {A.}~\bibnamefont
  {Puglisi}}, \bibinfo {author} {\bibfnamefont {L.}~\bibnamefont {Rondoni}},\
  and\ \bibinfo {author} {\bibfnamefont {A.}~\bibnamefont {Vulpiani}},\
  }\bibfield  {title} {\bibinfo {title} {Fluctuation–dissipation: Response
  theory in statistical physics},\ }\href@noop {} {\bibfield  {journal}
  {\bibinfo  {journal} {Phys. Rep.}\ }\textbf {\bibinfo {volume} {461}},\
  \bibinfo {pages} {111} (\bibinfo {year} {2008})}\BibitemShut {NoStop}%
\bibitem [{\citenamefont {Ishii}(1985)}]{ishii1985}%
  \BibitemOpen
  \bibfield  {author} {\bibinfo {author} {\bibfnamefont {T.}~\bibnamefont
  {Ishii}},\ }\bibfield  {title} {\bibinfo {title} {Theory of classical hopping
  conduction},\ }\href {https://doi.org/10.1143/PTP.73.1084} {\bibfield
  {journal} {\bibinfo  {journal} {Progress of Theoretical Physics}\ }\textbf
  {\bibinfo {volume} {73}},\ \bibinfo {pages} {1084} (\bibinfo {year}
  {1985})}\BibitemShut {NoStop}%
\bibitem [{\citenamefont {Seifert}(2008)}]{Seifert2008}%
  \BibitemOpen
  \bibfield  {author} {\bibinfo {author} {\bibfnamefont {U.}~\bibnamefont
  {Seifert}},\ }\bibfield  {title} {\bibinfo {title} {Stochastic
  thermodynamics: Principles and perspectives},\ }\href@noop {} {\bibfield
  {journal} {\bibinfo  {journal} {Eur. Phys. J. B}\ }\textbf {\bibinfo {volume}
  {64}},\ \bibinfo {pages} {423} (\bibinfo {year} {2008})}\BibitemShut
  {NoStop}%
\bibitem [{\citenamefont {Seifert}(2012)}]{Seifert2012}%
  \BibitemOpen
  \bibfield  {author} {\bibinfo {author} {\bibfnamefont {U.}~\bibnamefont
  {Seifert}},\ }\bibfield  {title} {\bibinfo {title} {Stochastic
  thermodynamics, fluctuation theorems and molecular machines},\ }\href@noop {}
  {\bibfield  {journal} {\bibinfo  {journal} {Rep. Prog. Phys.}\ }\textbf
  {\bibinfo {volume} {75}},\ \bibinfo {pages} {126001} (\bibinfo {year}
  {2012})}\BibitemShut {NoStop}%
\bibitem [{\citenamefont {Van~den Broeck}\ and\ \citenamefont
  {Esposito}(2015)}]{VandenBroeck2015}%
  \BibitemOpen
  \bibfield  {author} {\bibinfo {author} {\bibfnamefont {C.}~\bibnamefont
  {Van~den Broeck}}\ and\ \bibinfo {author} {\bibfnamefont {M.}~\bibnamefont
  {Esposito}},\ }\bibfield  {title} {\bibinfo {title} {Ensemble and trajectory
  thermodynamics: A brief introduction},\ }\href@noop {} {\bibfield  {journal}
  {\bibinfo  {journal} {Physica A}\ }\textbf {\bibinfo {volume} {418}},\
  \bibinfo {pages} {6} (\bibinfo {year} {2015})}\BibitemShut {NoStop}%
\bibitem [{\citenamefont {Peliti}\ and\ \citenamefont
  {Pigolotti}(2021)}]{Peliti2021}%
  \BibitemOpen
  \bibfield  {author} {\bibinfo {author} {\bibfnamefont {L.}~\bibnamefont
  {Peliti}}\ and\ \bibinfo {author} {\bibfnamefont {S.}~\bibnamefont
  {Pigolotti}},\ }\href@noop {} {\emph {\bibinfo {title} {Stochastic
  Thermodynamics: an Introduction}}}\ (\bibinfo  {publisher} {Princeton
  University Press},\ \bibinfo {address} {Princeton},\ \bibinfo {year}
  {2021})\BibitemShut {NoStop}%
\bibitem [{\citenamefont {Brandner}\ \emph {et~al.}(2015)\citenamefont
  {Brandner}, \citenamefont {Saito},\ and\ \citenamefont
  {Seifert}}]{brandner2015thermodynamics}%
  \BibitemOpen
  \bibfield  {author} {\bibinfo {author} {\bibfnamefont {K.}~\bibnamefont
  {Brandner}}, \bibinfo {author} {\bibfnamefont {K.}~\bibnamefont {Saito}},\
  and\ \bibinfo {author} {\bibfnamefont {U.}~\bibnamefont {Seifert}},\
  }\bibfield  {title} {\bibinfo {title} {Thermodynamics of micro-and
  nano-systems driven by periodic temperature variations},\ }\href@noop {}
  {\bibfield  {journal} {\bibinfo  {journal} {Physical review X}\ }\textbf
  {\bibinfo {volume} {5}},\ \bibinfo {pages} {031019} (\bibinfo {year}
  {2015})}\BibitemShut {NoStop}%
\bibitem [{\citenamefont {Proesmans}\ and\ \citenamefont {Van~den
  Broeck}(2015)}]{proesmans2015onsager}%
  \BibitemOpen
  \bibfield  {author} {\bibinfo {author} {\bibfnamefont {K.}~\bibnamefont
  {Proesmans}}\ and\ \bibinfo {author} {\bibfnamefont {C.}~\bibnamefont
  {Van~den Broeck}},\ }\bibfield  {title} {\bibinfo {title} {Onsager
  coefficients in periodically driven systems},\ }\href@noop {} {\bibfield
  {journal} {\bibinfo  {journal} {Physical review letters}\ }\textbf {\bibinfo
  {volume} {115}},\ \bibinfo {pages} {090601} (\bibinfo {year}
  {2015})}\BibitemShut {NoStop}%
\bibitem [{\citenamefont {Brandner}\ and\ \citenamefont
  {Seifert}(2016)}]{brandner2016periodic}%
  \BibitemOpen
  \bibfield  {author} {\bibinfo {author} {\bibfnamefont {K.}~\bibnamefont
  {Brandner}}\ and\ \bibinfo {author} {\bibfnamefont {U.}~\bibnamefont
  {Seifert}},\ }\bibfield  {title} {\bibinfo {title} {Periodic thermodynamics
  of open quantum systems},\ }\href@noop {} {\bibfield  {journal} {\bibinfo
  {journal} {Physical Review E}\ }\textbf {\bibinfo {volume} {93}},\ \bibinfo
  {pages} {062134} (\bibinfo {year} {2016})}\BibitemShut {NoStop}%
\bibitem [{\citenamefont {Proesmans}\ \emph {et~al.}(2016)\citenamefont
  {Proesmans}, \citenamefont {Cleuren},\ and\ \citenamefont {Van~den
  Broeck}}]{Proesmans2016}%
  \BibitemOpen
  \bibfield  {author} {\bibinfo {author} {\bibfnamefont {K.}~\bibnamefont
  {Proesmans}}, \bibinfo {author} {\bibfnamefont {B.}~\bibnamefont {Cleuren}},\
  and\ \bibinfo {author} {\bibfnamefont {C.}~\bibnamefont {Van~den Broeck}},\
  }\bibfield  {title} {\bibinfo {title} {Linear stochastic thermodynamics for
  periodically driven systems},\ }\href@noop {} {\bibfield  {journal} {\bibinfo
   {journal} {J. Stat. Mech.: Theor. Exp.}\ }\textbf {\bibinfo {volume}
  {2016}},\ \bibinfo {pages} {023202} (\bibinfo {year} {2016})}\BibitemShut
  {NoStop}%
\bibitem [{\citenamefont {Proesmans}\ and\ \citenamefont
  {Fiore}(2019)}]{Proesmans2019}%
  \BibitemOpen
  \bibfield  {author} {\bibinfo {author} {\bibfnamefont {K.}~\bibnamefont
  {Proesmans}}\ and\ \bibinfo {author} {\bibfnamefont {C.~E.}\ \bibnamefont
  {Fiore}},\ }\bibfield  {title} {\bibinfo {title} {General linear
  thermodynamics for periodically driven systems with multiple reservoirs},\
  }\href@noop {} {\bibfield  {journal} {\bibinfo  {journal} {Phys. Rev. E}\
  }\textbf {\bibinfo {volume} {100}},\ \bibinfo {pages} {022141} (\bibinfo
  {year} {2019})}\BibitemShut {NoStop}%
\bibitem [{\citenamefont {Cleuren}\ and\ \citenamefont
  {Proesmans}(2019)}]{Cleuren2019}%
  \BibitemOpen
  \bibfield  {author} {\bibinfo {author} {\bibfnamefont {B.}~\bibnamefont
  {Cleuren}}\ and\ \bibinfo {author} {\bibfnamefont {K.}~\bibnamefont
  {Proesmans}},\ }\bibfield  {title} {\bibinfo {title} {Stochastic impedance},\
  }\href@noop {} {\bibfield  {journal} {\bibinfo  {journal} {Physica A:
  Statistical Mechanics and its Applications}\ ,\ \bibinfo {pages} {122789}}
  (\bibinfo {year} {2019})}\BibitemShut {NoStop}%
\bibitem [{\citenamefont {Forastiere}\ \emph {et~al.}(2022)\citenamefont
  {Forastiere}, \citenamefont {Rao},\ and\ \citenamefont
  {Esposito}}]{Forastiere2022}%
  \BibitemOpen
  \bibfield  {author} {\bibinfo {author} {\bibfnamefont {D.}~\bibnamefont
  {Forastiere}}, \bibinfo {author} {\bibfnamefont {R.}~\bibnamefont {Rao}},\
  and\ \bibinfo {author} {\bibfnamefont {M.}~\bibnamefont {Esposito}},\
  }\bibfield  {title} {\bibinfo {title} {Linear stochastic thermodynamics},\
  }\href {https://doi.org/10.1088/1367-2630/ac836b} {\bibfield  {journal}
  {\bibinfo  {journal} {New Journal of Physics}\ }\textbf {\bibinfo {volume}
  {24}},\ \bibinfo {pages} {083021} (\bibinfo {year} {2022})}\BibitemShut
  {NoStop}%
\bibitem [{\citenamefont {{Van Kampen}}(2007)}]{vankampen}%
  \BibitemOpen
  \bibfield  {author} {\bibinfo {author} {\bibfnamefont {N.}~\bibnamefont {{Van
  Kampen}}},\ }\href@noop {} {\emph {\bibinfo {title} {Stochastic Processes in
  Physics and Chemistry}}},\ \bibinfo {edition} {3rd}\ ed.\ (\bibinfo
  {publisher} {Elsevier},\ \bibinfo {address} {Amsterdam},\ \bibinfo {year}
  {2007})\BibitemShut {NoStop}%
\bibitem [{\citenamefont {Schnakenberg}(1976)}]{schnakenberg1976}%
  \BibitemOpen
  \bibfield  {author} {\bibinfo {author} {\bibfnamefont {J.}~\bibnamefont
  {Schnakenberg}},\ }\bibfield  {title} {\bibinfo {title} {Network theory of
  microscopic and macroscopic behavior of master equation systems},\
  }\href@noop {} {\bibfield  {journal} {\bibinfo  {journal} {Rev. Mod. Phys.}\
  }\textbf {\bibinfo {volume} {48}},\ \bibinfo {pages} {571} (\bibinfo {year}
  {1976})}\BibitemShut {NoStop}%
\end{thebibliography}%

\appendix

\end{document}